\documentclass[numberedappendix,apj]{emulateapj} 
\usepackage{apjfonts} 
\usepackage{bm}

\newcommand{\refeqn}[1]{(\ref{#1})}

\shortauthors{Lange \& Page} 
\shorttitle{Expanding CMB Photosphere} 
\slugcomment{Submitted to The Astrophysical Journal} 
 
\begin{document} 
 
\title{Measuring the Expansion of the Universe Through 
Changes in the CMB Photosphere } 
 
\author{S. Lange \& L. Page } 
\affil{Princeton University, Department of Physics, Jadwin Hall,  
Washington Rd., Princeton, NJ, 08544}

\begin{abstract} 
 
The expansion of the universe may be observed in ``realtime'' by  
measuring changes in the patterns of the anisotropy in the CMB. 
As the universe ages, the surface of decoupling---or the  
CMB photosphere---moves away from us and samples a different gravitational  
landscape. The response of the CMB to this new landscape results in 
a different pattern than we observe today. The largest change occurs  
at $\ell\approx 900$. We show that with an array of detectors that we  
may envision having in a couple of decades, one can in principle measure  
the change in the anisotropy with two high precision measurements  
separated by a century.   
  
\end{abstract} 
 
  
\section{Introduction} 
 
Measurements of the anisotropy of the cosmic microwave background 
(CMB) provide a snapshot of the universe some 380,000 years after  
the big bang, when the primordial plasma decouples from the baryons. 
The power spectrum of the anisotropy is computed with a line of sight 
integration of the coupled Boltzmann equations between now and  
some time well before decoupling \cite{selzal} (hereafter paper 1).  
This formalism may be  
extended to consider future times when the decoupling surface is further  
away from us. We may think of this as investigating 
an expanding CMB photosphere. Of course we do not know the details of the  
gravitational landscape beyond the current photosphere 
and so we can only probe the future in a statistical sense. 
 
The conformal time of the decoupling surface in a flat universe 
is obtained from the Friedmann equation as follows: 
\begin{equation} 
\label{eq:tau} 
\tau(a)=\int_0^a\frac{da^\prime} 
{H_0\sqrt{\Omega_{r,0}+a^\prime\Omega_{m,0}+a^{\prime 4}\Omega_{\Lambda,0}}}, 
\end{equation} 
where $a$ is the scale factor, $H_0$ is the current Hubble parameter,  
$\Omega_{r,0}$ is the current radiation density relative to the  
critical density, $\Omega_{m,0}$ is the current matter density 
relative to the critical density, and  
$\Omega_{\Lambda,0}$ is the current dark energy density relative 
to the critical density. At all times $\Omega_r+\Omega_m+\Omega_\Lambda=1$. 
With parameters inspired by the Wilkinson Microwave Anisotropy Probe (WMAP) 
\cite{spergel07}, we take $\Omega_r=8.056\times10^{-5}$, $\Omega_m=0.257$,  
$\Omega_\Lambda=0.743$ and $H_0=72~{\rm km/s/Mpc}$. The current conformal 
distance  to the decoupling surface ($a=1$ in the above equation) 
is $c\tau=14373$ Mpc. 
 
\begin{deluxetable}{ccc} 
\tablecaption{Time slices and Scale Factors} 
\tablehead{ 
\colhead{Time slice} &  
\colhead{$c\tau$ } &  
\colhead{Scale factor}\\ 
\colhead{Years} &  
\colhead{Mpc } &  
\colhead{$a$}} 
\startdata 
0  &  14373 & 1.0000  \\ 
$1\times10^8$ & 14403   &   1.0074  \\  
$5\times10^8$ & 14523   &   1.0372  \\  
$1\times10^9$ & 14669   &   1.0754  \\  
$2\times10^9$ & 14944   &   1.1545  \\  
$4\times10^9$ & 15440   &   1.3255  \\  
$5\times10^9$ & 15663   &   1.4182  \\  
$6\times10^9$ & 15872   &   1.5163  \\  
$8\times10^9$ & 16251   &   1.7304  \\  
$1\times10^{10}$ & 16583   &   1.9714  \\  
$4\times10^{10}$ & 18657   &   13.340  \\  
$1\times10^{11}$ & 19011   &   601.95  \\  
\enddata 
\tablecomments{To find the physical time of any time slice 
add $13.66\times10^9$ years, the current age of the universe for  
our adopted parameters. Because $d\tau=dt/a$, the physical time is  
computed from equation~\ref{eq:tau} except with an additional factor 
of $1/a^\prime$ in the integrand. Note that the universe doubles  
in diameter when the physical time is 75\% greater than today 
and the conformal time is 15\% greater.}  
\end{deluxetable}

\section{Computing the Power Spectrum} 
 
To perform our study, we use CAMB \cite{lewis} which follows the 
algorithm in Seljak  and Zaldarriga's CMBFAST \cite{selzal}. We 
consider only scalar perturbations.  We modified the CAMB code so 
that it computes the following  
evolution function (eq. 13 in paper 1) 
 
\begin{equation} 
\label{eqn:los_9} 
\Delta_{T_\ell}(\tau_f, k)  =  \int_{0}^{\tau_f} S(\tau, k)  
j_\ell\left[k(\tau_f - \tau)\right]d\tau 
\end{equation} 
where $S$ is the source function, $j_\ell$ is a spherical Bessel function, 
$\tau$ is the conformal time, and $k$ is the comoving wavevector of the  
perturbation. The modification allows for arbitrary final $\tau$ 
(as opposed to fixing $\tau_f = \tau_0$). 
The source function is given by 
\begin{equation} 
\label{eqn:source_3} 
S(k, \tau) =  
e^{\kappa_f} \left[  
g \left(\Delta_{T_0} + \Psi \right) + 
\frac{\partial}{\partial\tau}\left( g \frac{v_b}{k} \right) + 
e^{-\kappa}\left( \dot \Phi + \dot \Psi \right) 
+ P(k, \tau)\right] 
\end{equation} 
where $P(k,\tau)$ represents polarization terms (see eq. 12 in paper 1) which  
we ignore for simplicity. Here 
$g(\tau) = -\dot \kappa e^{-\kappa}$ is the visibility function 
with $\kappa(\tau)$ the optical depth from $\tau_0$ (note that 
for $\tau_f > \tau_0$, $\kappa_f < 0$). Eq. \refeqn{eqn:source_3} 
differs from eq. 12 in paper 1 only 
in the overall $e^{\kappa_f}$ term which accounts for the changing 
optical depth. The first order term in the  
expansion of the Boltzmann equation, $\Delta_{T0}$  
(eq. 3a in paper 1), the potentials $\Psi$ and $\Phi$, 
the velocity of the baryons $v_b$, and the polarization terms 
are computed by CAMB and need no modification.  The primary modification 
to CAMB is to calculate $S$ out to $\tau_f$, rather than 
to $\tau_0$.\footnote{We  
accomplish this simply by setting CAMB's internal $\tau_0$ variable 
to the maximum $\tau_f$ in which we are interested.  This is valid because 
in the code the concept of the ``present'' is linked to $a=1$ (which we do 
not modify), rather than to $\tau=\tau_0$.} In addition, some of CAMB's
optimizations that led to sparse sampling at recent times were removed. 
 
\begin{figure}[hbt] 
\epsscale{1.1} 
\plotone{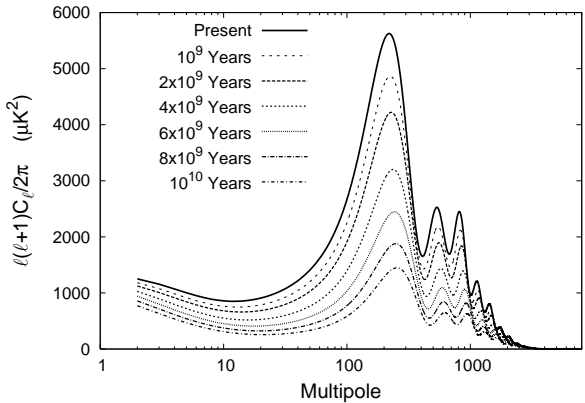} 
\figcaption[f1.eps]{The temperature angular power  
spectrum of the CMB at several  
representative time steps into the future. The scale factors and conformal 
distances are given in Table 1. The shift to the 
right of the first acoustic peak for some future time is  
$\ell_f=\ell_0(\tau_f-\tau_{dec})/(\tau_0-\tau_{dec})$ where  
$\tau_{dec}=286~{\rm Mpc}$ 
is the conformal time of decoupling, and $\ell_0=220$ 
the current $\ell$ of the acoustic peak. 
\label{fig:ps1} } 
\end{figure}

To compute the power spectrum at any time in the future we simply form 
the analog of eq 9 in paper 1: 
\begin{equation} 
\label{cmb:eqn:cl_8} 
C_\ell(\tau_f) = (4\pi)^2 \int k^2  P_\Psi(k) \left[   
\Delta_{T_{\ell}}(0,\tau_f, k)   
\Delta_{T_{\ell}}^*(0,\tau_f, k) \right]dk 
\end{equation} 
where $P_\Psi(k)$ is the initial power spectrum. Figure~\ref{fig:ps1} 
shows the power spectrum for future times. We see three major  
effects: (1) The power spectrum amplitude drops off due to the $1/a$  
scaling of the CMB temperature;\footnote{Note that Eqs. \refeqn{eqn:los_9} 
and \refeqn{cmb:eqn:cl_8} are dimensionless (since $\Delta_T$ is 
dimensionless by definition).  We 
give units to $C_\ell(\tau_f)$ by multiplying by $T_0^2/a^2(\tau_f)$, 
accounting for this effect.} 
(2) The features shift to smaller angular  
scales due to the recession of the surface of last scattering; (3)  
The low-$\ell$ tail becomes enhanced compared to the peak due to the  
integrated Sachs-Wolfe (ISW) effect caused by the shift to  
dark-energy dominance. The enhanced  ISW effect is not present in runs  
of the code without dark energy. In roughly $25\times10^9$ years, the ISW  
effect at $\ell=2$ will exceed the height of the first acoustic peak.  
 
\section{Maps of the future sky} 
 
To visualize the evolution of the CMB photosphere, we make maps of the sky 
for various physical time slices (see Table 1). For any particular map,  
which we take as a Gaussian random field, all the information is contained 
in the power  spectrum, $C_\ell=<a_{\ell m}a^\star_{\ell m}>$, where 
the $a_{\ell m}s$ are the  
coefficients of a spherical harmonic decomposition of the map. 
To generate a visualization, one draws $\ell+1$ random normal complex 
deviates, r, with variance $C_\ell$ to generate a set 
of $2\ell+1$ $a_{\ell m}$s that satisfy $a_{\ell -m}=a_{\ell m}^\star$. 
From these one forms the real valued 
$T(\theta,\phi)=\sum_{\ell m}a_{\ell m}Y_{\ell m}(\theta,\phi)$.   
 
A set of maps that shows the future evolution of the CMB 
will be correlated. To account for this, we compute the full  
covariance matrix  
\begin{equation} 
\label{eqn:c_l} 
C_\ell^{ij} = (4\pi)^2 \int k^2  P_\Psi(k) \left[   
\Delta_{T_{\ell}}(0,\tau^i, k)   
\Delta_{T_{\ell}}^*(0,\tau^j, k) \right]dk, 
\end{equation}  
where $\tau^i$ is the conformal time at any time in the future. 
Thus for a sequence of say $n$ maps, we would compute a $n\times n$ matrix 
for each $\ell$.  
 
\begin{figure}[hbt] 
\plotone{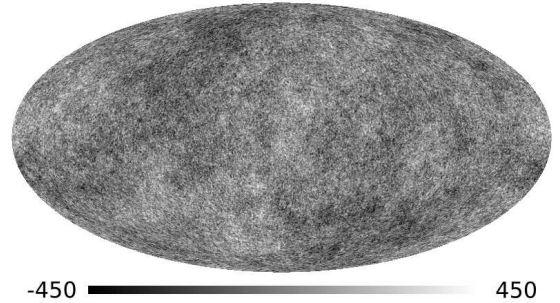} 
\figcaption{Present map for a set of random variables that  
follows the WMAP power spectrum. The units on the color bar are ${\rm \mu K}$. 
\label{fig:maps1} } 
\end{figure}

\begin{figure}[hbt] 
\plotone{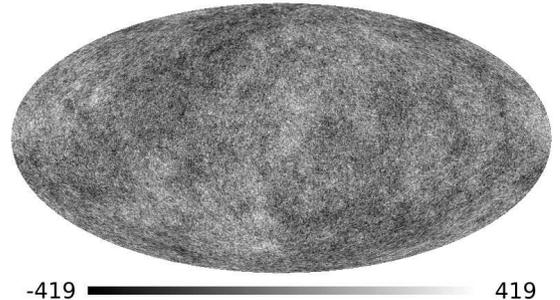} 
\figcaption{The same map as in Figure~\ref{fig:maps1} but now for 
a time slice $1\times10^9$ years in the future. Note that the color bar has  
been rescaled. The primary difference between the two maps is 
at small angular scales.  
\label{fig:maps2} } 
\end{figure}

\begin{figure}[hbt] 
\plotone{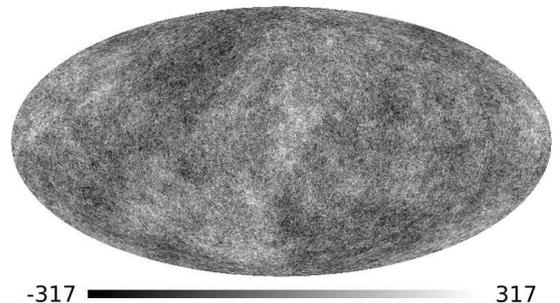} 
\figcaption{The same map as in Figure~\ref{fig:maps1} but now for 
a time slice $5\times10^9$ years in the future. Again, the color bar has  
been rescaled.  
\label{fig:maps3} } 
\end{figure}

\begin{figure}[hbt] 
\plotone{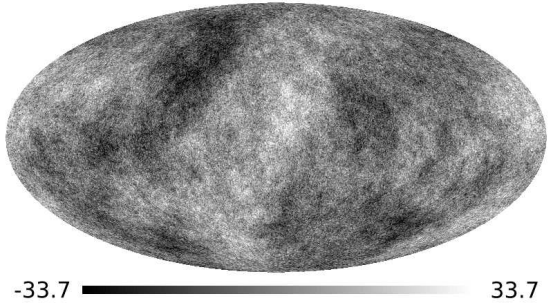} 
\figcaption{The same map as in Figure~\ref{fig:maps1} but now for 
a time slice $40\times10^9$ years in the future. The color bar has  
been rescaled. Note that the large angular scale fluctuations are more 
prominent relative to the smaller angular scales. 
\label{fig:maps4} } 
\end{figure}

We now extend the method given above to generate a set of correlated 
$a_{lm}$s. The first step is to decompose the covariance matrix as 
${\mathbf C}_\ell = {\mathbf M}_\ell{\mathbf D}_\ell{\mathbf M}_\ell^\star$ 
where ${\mathbf D}_\ell$ is a diagonal matrix. We then compute 
${\mathbf a}_{\ell m}={\mathbf M}_\ell[\sqrt{ {\mathbf D}_\ell }]{\mathbf r}$ 
where ${\mathbf r}$ is a vector of complex random deviates. 
This $\mathbf{a}_{\ell 
m}$ has the covariance matrix in eq.~\ref{eqn:c_l}.  
Figures~\ref{fig:maps1}-\ref{fig:maps4} show a set of four time 
slices starting 
with a random full-sky map that follows the WMAP parameters.\footnote{One 
could start with the WMAP sky though we have not done this.} One can see 
that most of the change occurs on small angular scales where the  
photosphere more quickly samples different potential wells as it expands. 
One also sees that as time progresses large angular scale fluctuations 
become more prominent as dark energy dominates the expansion. 
 
 
\begin{figure}[ht] 
\epsscale{1.1} 
\plotone{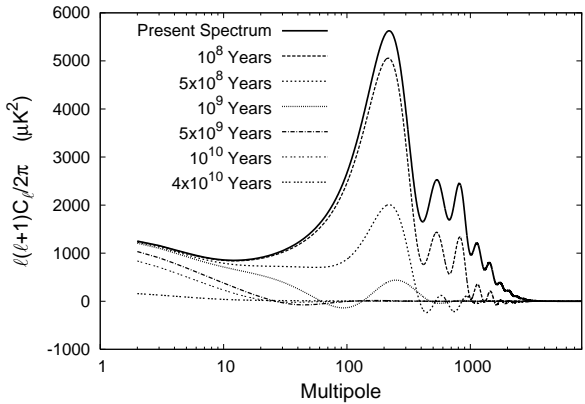} 
\caption{The $C_\ell^{0j}$ element of the covariance matrix for several  
time slices. The power spectrum at $\tau_0$ is also plotted for comparison.
Some of the decrease in the covariance is attributable to the decrease 
in the CMB temperature. For example, in $10^9$ years the power at 
$\ell=200$ decreases to 36\% the current value but 16\% of the 
decrease arrises because $T=T_0/a$.    
\label{fig:covariance_comp} } 
\end{figure}

Figure~\ref{fig:covariance_comp}  
shows elements of the covariance matrix as a function of time. 
As expected, nearby time slices are strongly correlated. As time  
progresses the covariance between current and future 
time slices at small angular scales disappears first, and then later the 
covariance at large angles decreases.  The time it takes for the scale 
factor to double, $\approx 10^{10}$y, gives a characteristic time for 
the future sky to decorrelate with what we observe today. 
 
\begin{figure}[ht] 
\epsscale{1.1} 
\plotone{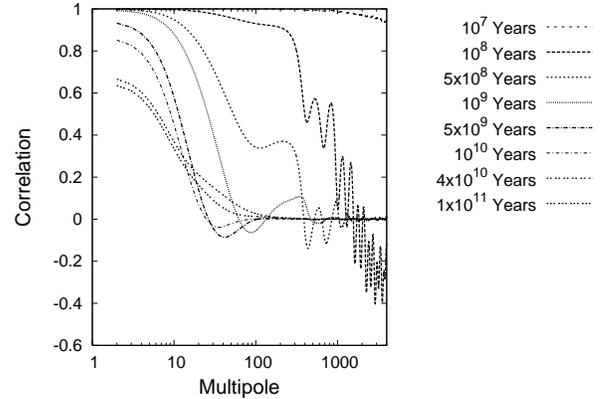} 
\caption{The correlation of several time slices with the present.   
\label{fig:correlation_comp} } 
\end{figure}

Figure~\ref{fig:correlation_comp} 
gives the correlation of several time steps with the present.  In this
plot, the decrease in temperature has been scaled out so that features
in the anisotropy may be compared directly. At high $\ell$, features in 
the sky at late times are uncorrelated with those at present. However, at 
low $\ell$, the late-time ISW features remain correlated for $>100$ billion 
years, indicating that these are very long lived structures.
 
\section{Measuring the Change in the CMB} 
 
Measuring the difference between two high precision maps of the  
anisotropy taken a century apart offers, in principle, a way to  
directly observe the  
expansion of the universe. Unlike a measurement of the temperature  
of the CMB, the difference between two maps is moderately insensitive to  
calibration. Rather, it is a change in spatial structure that is observed. 
Thus one needs a well understood pointing solution which is technically 
straightforward to achieve. 
 
\begin{figure}[ht] 
\epsscale{1.05} 
\plotone{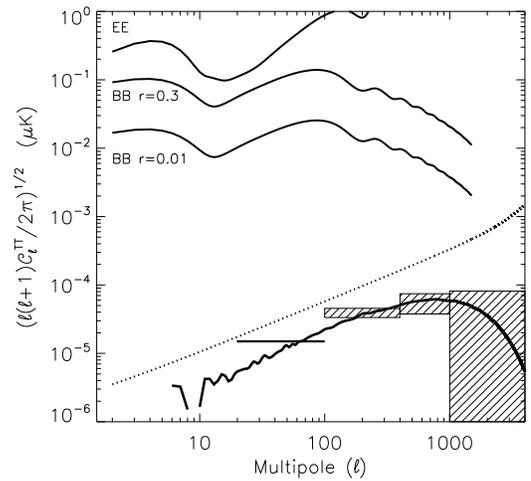} 
\figcaption[f8.eps]{The top curves shows the current CMB EE 
polarization spectrum for the parameters in this paper and 
a current optical depth of 0.1. The two curves below that 
show the BB spectrum for tensor to scalar ratios of $r=0.3$ and 
$r=0.01$. The solid curve at the bottom is the power spectrum 
of the difference between two CMB anisotropy maps made 100 years 
apart. The dotted line is Equation 6 and the hashed boxes show
the errors after binning in $\ell$.   
\label{fig:100yrs} } 
\end{figure}

Figure~\ref{fig:100yrs} shows the power spectrum of the  
difference between two maps taken 100 years apart. We use the formalism  
in Knox (1995) to compute the experimental uncertainty. Since there is  
just one sky realization, cosmic variance is ignored and the  
uncertainty per $\ell$ is 
\begin{equation} 
\Delta C_\ell= \sqrt{\frac{2}{f_{sky}(2\ell+1)}} 
\sigma^2\Omega_{pix}\exp(\ell^2\theta_{1/2}^2/8\ln 2), 
\end{equation} 
where $f_{sky}$ is the fraction of sky covered, $\sigma$ is the  
uncertainty per sky pixel of solid angle $\Omega_{pix}$, and $\theta_{1/2}$ 
is the FWHM of the beam profile. 
For the uncertainty bands shown in Figure~\ref{fig:100yrs}, we imagine 
an array of 3000x3000 detectors at 150~GHz each with a sensitivity of 
40~mK\,s$^{1/2}$ \cite{weissreport} observing the sky with $0.86^\prime$  
angular resolution. The observations would last 4 years,   
cover 75\% of the sky, and would have to be done from a satellite.  
The only element not already demonstrated is large array. Currently,  
arrays of 32x32 detectors are being built. 
 
The fundamental limit to such a measurement is likely to be variable 
point sources and variable foreground emission. Though in principle these  
can be identified and removed spectrally, this capability would add  
complexity to the ``simple'' scheme outlined above. For comparison, 
there are currently experiments being designed with the sensitivity to 
measure the  
$r=0.01$ B-mode polarization. The improvement to go from these planned  
missions to measuring the signal we describe is on the order of  
the improvement between observations of the 1980s and the current  
observations. 
 
\acknowledgments 
 
We gratefully acknowledge discussions with Mike Nolta, Uros Seljak and  
Paul Steinhardt. 
Uros provided a key insight for computing and manipulating the covariance 
matrix in equation~\ref{eqn:c_l}. In the course of this work, we learned  
that Adam Moss, Douglas Scott, and Jim Zibin were working on a more extensive  
version of a related calculation.  
A movie of the expanding photoshpere is available at 
http://phy-page-g5.princeton.edu/$\sim$page. 
This paper is based on Stuart Lange's senior  
thesis \cite{slange} and made extensive use of the publicly 
available HEALPix and CAMB software packages.  
The research was supported by NASA award  
LTSA03-000-0090 and by NSF Grant No. 0355328.

\clearpage

\end{document}